\documentclass[aps,preprint,nofootinbib,superscriptaddress]{revtex4-1}

\usepackage[utf8]{inputenc}
\usepackage[T1]{fontenc}
\usepackage{amsmath,amssymb}
\usepackage{graphicx}
\usepackage{xcolor}
\usepackage{hyperref}
\usepackage{slashed}
\usepackage{cleveref}
\newcommand{\ee}{\end{equation}}
\newcommand{\bb}{\begin{equation}}
\newcommand{\eqb}{\begin{eqnarray}}
\newcommand{\eqf}{\end{eqnarray}}

\begin{document}

\title{Infrared Quantum Electrodynamics and the Rayleigh-Jeans Physics}

\author{J.~Gamboa}
\affiliation{Departamento de Física, Universidad de Santiago de Chile, Santiago, Chile}
\email{jorge.gamboa@usach.cl}
\author{N.~Tapia-Arellano}
\affiliation{Department of Physics and Astronomy, Agnes Scott College, Decatur, GA. 30030, USA}
\email{narellano@agnesscott.edu}

\begin{abstract}
Infrared quantum electrodynamics (IR--QED) acquires a natural geometric 
interpretation once soft photons are described as adiabatically transported 
electron--photon clouds.  
Within this framework, the relevant infrared structure is encoded in a functional 
Berry phase associated with the space of gauge connections, and the corresponding 
Berry corrections modify the Rayleigh--Jeans spectrum.  
The infrared scaling symmetry of the Rayleigh--Jeans law leads to a simple 
renormalization--group equation whose solution determines the frequency dependence 
of an effective factor $F_{\rm eff}(\omega)$ controlling the strength of the 
electron--photon cloud dressing.  
As a result, the energy density of the cosmic microwave background (CMB) 
receives a Berry-induced correction that scales as a power law and produces a 
frequency-dependent temperature excess in the radio domain.  
Although the exponent $\gamma$ governing this scaling behaviour is not fixed 
internally by the present formulation of IR--QED and must instead be determined 
phenomenologically, the existence and structure of the excess are genuine 
predictions of the theory.  
Remarkably, the resulting expression is extremely simple and naturally aligns 
with the deviations suggested by the ARCADE~2  data.  
Taken together, these results indicate that Berry phases in IR--QED may lead to 
observable consequences in the low-frequency tail of the CMB spectrum.
\end{abstract}

\maketitle
\section{Introduction}
The infrared sector of quantum electrodynamics (IR-QED) remains one of the most subtle and conceptually challenging domains in quantum field theory. Over the decades, hundreds of papers have been devoted to understanding the physical origin and implications of infrared (IR) divergences.

The problem was first addressed by Bloch and Nordsieck~\cite{BN}, and later by other authors~\cite{KI,LN,YFS,Weinberg}, who proposed that in order to compute scattering amplitudes in processes involving soft photons, one must coherently sum over all initial and final states containing such photons, assuming that the detector resolution energy provides a natural infrared cutoff. Under this prescription, all diagrams containing IR divergences cancel order by order, leading to finite results.

Although the original formulation did not employ the modern language of regularization or renormalization-group ideas, the Bloch--Nordsieck approach \cite{BN} can in fact be viewed as an early contribution in this direction, since the mere introduction of a finite resolution scale already produces physically finite results, while still leaving open the issue of how the cutoff is fixed and how the amplitudes evolve with it.

The technical and conceptual difficulty, however, lies in understanding the physical origin of these divergences. This point was reexamined by Chung \cite{chung} and Kibble \cite{kibble1,kibble2,kibble3,kibble4} on one side, and by Kulish and Faddeev on the other \cite{KF}, who introduced the idea of \emph{dressing the asymptotic states of QED}. Technically, this means that in the infrared regime the in/out states of QED should not be regarded as elements of the Fock space, but rather as \emph{dressed states} of the form
\begin{equation}
|{\rm in}\rangle \;\longrightarrow\; {\cal U}[A,e]\,|{\rm in}\rangle,
\qquad
|{\rm out}\rangle \;\longrightarrow\; {\cal U}[A,e]\,|{\rm out}\rangle ,
\end{equation}
where the operator
\begin{equation}
{\cal U}[A,e] =
\exp\!\Big[i e \!\int d^3x\, f_\mu(A;x)\, a^\mu(x)\Big]
\label{KF}
\end{equation}
represents the Kulish--Faddeev (KF) dressing.
This operator generates a coherent cloud of soft photons surrounding the electron, producing what can be called an \emph{electron--photon cloud}. The dressed state thus constructed, retains its spin-$1/2$ character, since the fermionic field carries the spin degrees of freedom, while the coherent photon cloud provides a gauge dressing without modifying the spin content.

In recent works~\cite{JG1,JG2,JG3} we have developed an \emph{adiabatic formulation of IR-QED} that incorporates Berry’s ideas into the functional formalism of quantum field theory. Within this framework, the Kulish--Faddeev dressed states appear as \emph{adiabatically transported states}, and the accumulated phase defines a \emph{functional Berry phase}. We have shown that these states not only preserve the fermionic nature of the electron, but also that the KF dressing becomes quantized, so that IR-QED acquires a topologically protected structure, formally analogous to certain systems in condensed-matter physics, and conceptually related to the long-range quantum correlations often discussed in modern entanglement-based approaches to gauge theories \cite{sorella} (for a mathematical, discussion, see \cite{araki, araki1}).

It is important to stress, however, that the topological protection in IR-QED does not originate from the physical spacetime or energy-band topology, but rather emerges from the \emph{functional space of gauge connections}, $\mathcal{A}/\mathcal{G}$. In this sense, the topological properties of IR-QED are manifestations of the functional geometry of the gauge-field configuration space.

In this paper, we present a study of IR-QED,
streamlining the results obtained in~\cite{JG1,JG2,JG3} with the aim of discussing
blackbody radiation in the presence of soft photons.
There are several motivations for pursuing the analysis in this direction:
(a) in IR-QED, a soft photon is not a genuinely free excitation but an
electron--photon cloud, which is the physically relevant object in the infrared regime;
(b) although blackbody radiation is a well-established result and lies at the
foundations of modern quantum theory, reliable experimental measurements in the
low-frequency domain are technically difficult, and existing data still exhibit
significant uncertainties;
(c) cosmology provides an ideal setting for these considerations: the infrared
sector---in cosmological terms, the radio-frequency domain---is particularly
challenging to probe, and recent results such as those reported by the ARCADE-2
experiment, which indicate a temperature excess in the 3--10~GHz range, suggest
deviations that are consistent with the need to employ an infrared-sensitive
framework such as IR-QED, in which soft-photon clouds are genuine dynamical
degrees of freedom.

The article is organized as follows. In Sec.~II we present a discussion and a 
simplified formulation of the results developed in~\cite{JG1,JG2,JG3}, from which 
we derive a general expression for the Berry phase in IR-QED.  
In Sec.~III we study the Planck distribution for electron--photon clouds and show 
how Berry corrections naturally emerge in the radio-frequency sector.  
In Sec.~IV we confront our theoretical predictions with the existing observational 
data from ARCADE~2, and we show that our framework can account 
for the observed temperature excess.

\section{Discussing Infrared Quantum Electrodynamics}

In this section we derive the infrared formulation IR-QED 
and show how the interpretation of electron–photon clouds naturally emerges in this context. 
Within the same framework we also recover the usual Planck distribution 
as the free (Gaussian) contribution of the Maxwell sector, 
while identifying the infrared corrections that originate from the fermionic determinant.

To this end, let us start from the generating functional
\eqb
Z &=& \int [{\cal D}A]\,{\cal D}\bar\psi\,{\cal D}\psi~
\exp\!\left[i\int d^4x\left(
-\frac{1}{4}\,F_{\mu\nu}F^{\mu\nu}
+\bar\psi\,(i\slashed{D}[A]-m)\,\psi\right)\right] \nonumber\\[4pt]
&=& \int [{\cal D}A]~
\exp\!\left[i\int d^4x\,
\left(-\tfrac{1}{4}\,F_{\mu\nu}F^{\mu\nu}\right)\right]
\det\!\big(i\slashed{D}[A]-m\big),
\label{FG}
\eqf
where $[{\cal D}A]$ denotes the Faddeev--Popov measure.%
\footnote{Here ``tr'' denotes the trace over Dirac indices, 
while ``Tr'' refers to the functional trace over mode space.}

\vspace{2mm}
To evaluate the fermionic determinant in the adiabatic approximation, 
we assume that the background gauge field varies slowly with the Minkowski time variable $t$.
For each fixed $t$, we solve the instantaneous Dirac eigenvalue problem,
\begin{equation}
H_D(t)\,\varphi_m({\bf x};t) = E_m(t)\,\varphi_m({\bf x};t),
\end{equation}
where $H_D(t)= -i\gamma^0 \boldsymbol{\gamma}\!\cdot\!(\nabla - i\mathbf{A})$, 
and the set $\{\varphi_m\}$ forms an orthonormal basis.  
For massless fermions, the spectrum is symmetric under
$E_m \leftrightarrow -E_m$, so the net dynamical phase cancels between
conjugate pairs of modes.

The fermionic fields are expanded as
\begin{align}
\psi(t,{\bf x}) &= \sum_m a_m(t)\,\varphi_m({\bf x};t),\\
\bar{\psi}(t,{\bf x}) &= \sum_m \bar{a}_m(t)\,\varphi_m^\dagger({\bf x};t).
\end{align}
Substituting into the action gives%
\footnote{This procedure was first developed in Ref.~\cite{CGL}.}
\begin{equation}
S_F = \int dt\,\sum_{m,n}
\bar{a}_m(t)\Big[
  i\delta_{mn}\partial_t
  - i\mathcal{A}_{mn}(t)
  - E_m(t)\delta_{mn}
\Big]a_n(t),
\end{equation}
with the (anti-Hermitian) Berry connection
\begin{equation}
\mathcal{A}_{mn}(t)
   = i\,\langle \varphi_m | \partial_t | \varphi_n \rangle.
\end{equation}

In the chiral (massless) limit, the spectrum becomes symmetric under 
$E_m \leftrightarrow -E_m$, and the Berry connection acquires a 
non-Abelian structure within the degenerate subspaces of positive 
and negative energy modes, encoding the purely geometric evolution 
of the fermionic vacuum.

Performing the Grassmann path integral over the coefficients $a_m$ yields
\begin{equation}
\det(i\partial_t - E - i\mathcal{A})
   = \exp\!\Big[i\,\mathrm{Tr}\,
   \ln(i\partial_t - E - i\mathcal{A})\Big].
\end{equation}
In the adiabatic limit, where $\mathcal{A}$ varies slowly compared to
the instantaneous spectrum $E_m(t)$, the determinant reduces to
\begin{equation}
\det(i\slashed{D})
   \;\approx\;
   \exp\!\left[-i\int dt\, \sum_m E_m(t)\right]\,
   \mathrm{Tr}\,
   \exp\!\left(i\oint_C \mathcal{A}\right).
\end{equation}
For massless fermions\footnote{In the infrared regime the electron is never truly free, as it is 
always accompanied by a coherent cloud of soft photons. 
Nevertheless, in the adiabatic treatment we may formally take the 
massless limit ($m\!\to\!0$) to isolate the purely geometric part of 
the fermionic determinant. This limit does not imply that the physical 
electron is massless, but rather that the dynamical phase cancels 
between $\pm E_m$ pairs, leaving only the infrared holonomy that 
characterizes the dressed electron state.},
the first (dynamical) exponential cancels between
$\pm E_m$ pairs, while the second term—representing the holonomy of the
Berry connection—survives as the \emph{topological} contribution.%
\footnote{Henceforth we shall set the dynamical phase $E_m$ to zero.}

In other words,
\begin{equation}
   \det(i\slashed{D})
   \;\approx\;
   \mathrm{Tr}\,
   \exp\!\left(i\oint_C \mathcal{A}\right).
\end{equation}

This geometric phase reflects the parallel transport of the degenerate
eigenspaces and constitutes the infrared holonomy that defines the 
electron–photon cloud, i.e. the physical dressing responsible for the 
nontrivial IR structure of QED.

After integrating out the fermions, the generating functional takes the form
\[
Z = \int [\mathcal{D}A]\,
\exp\!\left\{ i S_{\text{Maxwell}}[A] + i S_{\text{IR}}[A]\right\},
\]
where $S_{\text{Maxwell}}[A]$ is quadratic in $A_\mu$ and 
$S_{\text{IR}}[A]$ encodes the infrared Berry phase structure obtained 
from the fermionic determinant.
If $S_{\text{IR}}[A]$ were absent, the Maxwell sector would reduce to a 
Gaussian functional integral, contributing only an overall normalization 
to the vacuum-to-vacuum amplitude (once the free-photon partition 
function is factored out). 
In the present context, however, we keep the quadratic Maxwell term, 
since it is precisely this piece that reproduces the standard Planck 
distribution for free photons. 
The new infrared contribution $S_{\text{IR}}[A]$ introduces corrections 
that encode the presence of dressed fermionic states.

It is crucial to stress that the infrared sector of IR-QED does not 
describe a gas of free bosonic photons. 
The physically relevant degrees of freedom are the electron–photon clouds, 
which remain fermionic in nature. 
The spin and statistics are carried by the underlying electron, while 
the soft-photon dressing provides a gauge-invariant completion of the 
state. Consequently, the quantization is not expressed in terms of photon 
creation and annihilation operators $a^\dagger,a$, but rather through 
quantized fluxes of the Berry connection,
\[
\Phi_n = \oint_C \mathcal{A} = n\,\Phi_0,
\]
which label distinct infrared sectors. 
Energy levels in this regime are therefore determined by the topological 
flux $\Phi_n$ instead of the bosonic occupation number. 
This distinction clarifies why IR-QED, although consistent with the 
Planck law in its free limit, represents a fundamentally different 
and topologically organized phase of QED.

\section{Planck distribution for IR-QED clouds and its Rayleigh--Jeans limit}
\label{subsec:Planck_clouds}

In the infrared regime of quantum electrodynamics (IR-QED), the physically relevant excitations are not bare photons but fermionic electron--photon clouds. 
After implementing the adiabatic approximation and incorporating the quantized Berry connection, the soft sector can be described effectively in terms of a discrete set of collective cloud modes. 
Each mode is characterized by a soft momentum $\mathbf{k}$ and by a quantized topological (Berry) sector $a$, which carries a definite geometric phase. 
In this subsection we derive the Planck distribution as a \emph{distribution of clouds} and then extract the Rayleigh--Jeans (RJ) limit, showing explicitly how the geometric corrections of the Berry connection generate small but structured infrared deviations.

\subsection{Effective spectrum of clouds in IR-QED}

The infrared Hamiltonian takes the form
\begin{equation}
H_{\rm IR}
=
\sum_{\mathbf{k},a}
\varepsilon_{\mathbf{k},a}\,
N_{\mathbf{k},a}
\;+\;
E_0,
\label{eq:H_IR_clouds}
\end{equation}
where
\begin{itemize}
    \item $\mathbf{k}$ labels the soft momentum modes,
    \item $a$ labels the quantized Berry (topological) sector,
    \item $N_{\mathbf{k},a}$ is the cloud occupation operator,
    \item and $E_0$ denotes the vacuum contribution.
\end{itemize}
The single–cloud energy in the mode $(\mathbf{k},a)$ is
\begin{equation}
\varepsilon_{\mathbf{k},a}
=
\hbar\omega_{\mathbf{k}}\,F_a ,
\label{eq:eps_cloud_def}
\end{equation}
with $\omega_{\mathbf{k}}=c|\mathbf{k}|$. 
The dimensionless factor $F_a$ arises from the Berry phase accumulated by the 
cloud in the adiabatic transport around its cycle in sector $a$. 
Explicitly, $F_a$ is the exponential of the Berry holonomy,
\[
F_a = \exp\!\Big[\,\oint_{\mathcal{C}_a} \mathcal{A}_\lambda\,dx^\lambda
        + \Delta\Phi_a^{\,(2)} + \cdots \Big],
\]
where the first term encodes the leading Berry connection of IR–QED and the
subsequent terms $\Delta\Phi_a^{\,(2)},\ldots$ represent higher–order
adiabatic corrections to the phase \cite{Berry2}.   
Thus $F_a$ is finite, sector–dependent, and fixed by the geometric 
(adiabatic–Berry) structure of the infrared theory.

Although the fundamental constituents of each cloud are fermionic, the collective infrared excitations behave effectively as bosonic modes. 
This follows from integrating out the fast fermionic degrees of freedom: the resulting infrared functional determinant produces a quadratic effective action for the soft gauge sector, whose normal modes satisfy Bose--Einstein statistics. 
{In other words, the Bose--Einstein distribution arises at the level of the effective infrared theory: after integrating out fermions in the adiabatic regime, the cloud sector is governed by a quadratic (Gaussian) action, so that the normal modes behave as harmonic oscillators. The integer occupation number therefore counts quanta of the collective cloud mode, not of a bare photon field.}
Hence, the integer $n_{\mathbf{k},a}=0,1,2,\ldots$ counts quanta of the coherent \emph{cloud field}, not of the elementary photon field. 
The effective Hamiltonian therefore describes a set of harmonic modes labeled by $(\mathbf{k},a)$, with energies
\begin{equation}
E_{n,\mathbf{k},a}
=
n\,\varepsilon_{\mathbf{k},a}
=
n\,\hbar\omega_{\mathbf{k}}\,F_a.
\label{eq:En_cloud}
\end{equation}

\subsection{Planck distribution for clouds}

Coupling the IR-QED cloud sector to a thermal bath at temperature $T=1/\beta$, the partition function factorizes into independent modes,
\begin{equation}
Z_{\rm IR}
=
\prod_{\mathbf{k},a} Z_{\mathbf{k},a},
\qquad
Z_{\mathbf{k},a}
=
\mathrm{Tr}\, e^{-\beta H_{\mathbf{k},a}^{\rm eff}},
\label{eq:ZIR_factorization}
\end{equation}
with
\begin{equation}
H_{\mathbf{k},a}^{\rm eff}
=
\varepsilon_{\mathbf{k},a}\,
N_{\mathbf{k},a}.
\label{eq:H_eff_mode}
\end{equation}
Since $N_{\mathbf{k},a}$ counts bosonic cloud quanta, the mode partition function is
\begin{equation}
Z_{\mathbf{k},a}
=
\sum_{n=0}^{\infty} e^{-\beta n\varepsilon_{\mathbf{k},a}}
=
\frac{1}{1-e^{-\beta\varepsilon_{\mathbf{k},a}}}.
\label{eq:Zka_BE}
\end{equation}
The mean occupation number is therefore
\begin{equation}
\langle N_{\mathbf{k},a}\rangle
=
\frac{1}{e^{\beta\varepsilon_{\mathbf{k},a}}-1}
=
\frac{1}{e^{\beta\hbar\omega_{\mathbf{k}}F_a}-1},
\label{eq:occupation_cloud_BE}
\end{equation}
which has the form of a Bose--Einstein distribution but now refers to collective cloud excitations rather than bare photons. 
The average energy per mode is
\begin{equation}
\langle E_{\mathbf{k},a}\rangle
=
\frac{\hbar\omega_{\mathbf{k}}F_a}{e^{\beta\hbar\omega_{\mathbf{k}}F_a}-1}.
\label{eq:Energy_mode_cloud}
\end{equation}

In the continuum limit, summing over all Berry sectors gives the energy density per unit frequency,
\begin{equation}
u_{\rm cloud}(\omega,T)\,d\omega
=
\frac{\omega^2}{2\pi^2c^3}\,d\omega
\sum_a
\hbar\omega F_a
\frac{1}{e^{\beta\hbar\omega F_a}-1}.
\label{eq:u_cloud_general}
\end{equation}
Assuming that a dominant sector $a=a_\star$ governs the infrared regime, or that all sectors can be encoded in an effective factor $F_{\rm eff}(\omega)$ defined as
\begin{equation}
F_{\rm eff}(\omega)
\equiv
\frac{\sum_a F_a\,\big[\exp(\beta\hbar\omega F_a)-1\big]^{-1}}
{\sum_a \big[\exp(\beta\hbar\omega F_a)-1\big]^{-1}},
\label{eq:F_eff_def}
\end{equation}
one obtains the compact expression
\begin{equation}
u_{\rm cloud}(\omega,T)
=
\frac{\hbar\omega^3}{\pi^2c^3}\,
\frac{F_{\rm eff}(\omega)}{e^{\beta\hbar\omega F_{\rm eff}(\omega)}-1}.
\label{eq:Planck_clouds}
\end{equation}
{At this level, Berry-phase effects enter through the effective cloud energy $\varepsilon_{\mathbf{k}}=\hbar\omega F_{\rm eff}(\omega)$, while the standard phase-space density of electromagnetic modes is preserved. Any infrared modification induced by the Berry holonomy is therefore encoded in the factor $F_{\rm eff}(\omega)$ rather than in an explicit change of the mode-counting factor.}
Equation~\eqref{eq:Planck_clouds} represents the \emph{Planck law for IR-QED clouds}: it retains the familiar functional form of black-body radiation, but the role of $\hbar\omega$ is now played by the cloud energy $\varepsilon_{\mathbf{k},a}$, which embeds the quantized Berry contribution. 
In the limit $F_{\rm eff}(\omega)\!\to\!1$, the usual Planck spectrum of bare photons is recovered.

\subsection{Rayleigh--Jeans limit and geometric corrections}

In the low-frequency limit define $x(\omega)=\beta\hbar\omega F_{\rm eff}(\omega)$.  
For $x\ll1$,
\begin{equation}
\frac{1}{e^{x}-1}
=
\frac{1}{x}-\frac{1}{2}+\frac{x}{12}-\frac{x^3}{720}+\cdots,
\label{eq:BE_expansion}
\end{equation}
which leads to
\begin{equation}
\langle E_{\mathbf{k}}\rangle
\simeq
k_BT
-\frac{1}{2}\hbar\omega F_{\rm eff}(\omega)
+\frac{(\beta\hbar\omega F_{\rm eff}(\omega))^2}{12}\,k_BT
+\cdots.
\label{eq:E_mode_expanded}
\end{equation}
The leading term reproduces equipartition and gives
\begin{equation}
u_{\rm RJ}^{\rm cloud}(\omega,T)
=
\frac{\omega^2}{\pi^2c^3}\,k_BT,
\label{eq:RJ_cloud}
\end{equation}
the standard Rayleigh--Jeans law. 
The Berry dependence cancels at this order but reappears in the subleading corrections,
\begin{equation}
u_{\rm cloud}(\omega,T)
=
u_{\rm RJ}^{\rm cloud}(\omega,T)
+\Delta u_{\rm Berry}(\omega,T),
\label{eq:u_cloud_RJ_plus_corr}
\end{equation}
with
\begin{equation}
\Delta u_{\rm Berry}(\omega,T)
=
-\frac{\hbar\omega^3}{2\pi^2c^3}\,F_{\rm eff}(\omega)
+\frac{\beta(\hbar\omega)^2\omega^2}{12\pi^2c^3}\,F_{\rm eff}^2(\omega)
+\cdots.
\label{eq:Delta_u_Berry}
\end{equation}

In the infrared regime, the effective factor $F_{\text{eff}}(\omega)$
characterizes the response of the vacuum---or equivalently, of the adiabatic medium---
to soft excitations of frequency $\omega$.
Although its exact functional form cannot be determined without solving the full dynamics,
its asymptotic behaviour can be constrained by general renormalization-group
considerations.

To determine this behaviour, let us consider in general terms
a renormalized quantity $F(\omega,\mu,g)$ that depends on the physical energy scale $\omega$,
the renormalization scale $\mu$, and a coupling $g$.
Under these conditions, the Callan--Symanzik equation reads
\begin{equation}
\left(
\mu\,\frac{\partial}{\partial\mu}
+ \beta(g)\,\frac{\partial}{\partial g}
+ \gamma_F(g)
\right)
F(\omega,\mu,g) = 0,
\end{equation}
where $\beta(g)=\mu\,\dfrac{\partial g}{\partial\mu}$ is the beta function
and $\gamma_F(g)$ is the anomalous dimension associated with $F$.

By dimensional analysis, the function $F$ can depend on $\omega$ and $\mu$
only through the dimensionless ratio $\omega/\mu$,
so that
\[
F(\omega,\mu,g) = \mathcal{F}\!\left(\frac{\omega}{\mu},\,g(\mu)\right).
\]
Upon varying the renormalization scale, one obtains an equivalent
Callan--Symanzik equation written in terms of the physical frequency:
\begin{equation}
\left(
\omega\,\frac{\partial}{\partial\omega}
- \beta(g)\,\frac{\partial}{\partial g}
- \gamma_F(g)
\right)
\mathcal{F}(\omega,g) = 0.
\label{eq:CS_freq}
\end{equation}
{Here the condition $\beta(g)\simeq 0$ should be understood as an \emph{effective infrared scaling assumption} for the cloud--dressing factor $F_{\rm eff}$, rather than as a statement about a fundamental conformal fixed point of microscopic QED. In the Rayleigh--Jeans window, the adiabatic infrared sector is dominated by slowly varying collective modes, so that the residual running of the microscopic coupling becomes subleading compared to the $\omega$--scaling of the Berry-induced dressing.}

\noindent

In the deep infrared (Rayleigh--Jeans) window relevant for the present
discussion, it is natural to consider an \emph{approximate scaling regime} in
which the running relevant for $F_{\rm eff}$ becomes slow, so that
$\beta(g(\omega))\,\partial_g F_{\rm eff}$ is subleading compared to
$\omega\,\partial_\omega F_{\rm eff}$, while the anomalous dimension approaches
an approximately constant value,
\[
\gamma_F\bigl(g(\omega)\bigr)\;\longrightarrow\;\gamma \, .
\]
Under these conditions, Eq.~(\ref{eq:CS_freq}) reduces asymptotically to the
linearised form

\[
\gamma_F(g) \xrightarrow[\omega\to 0]{} \gamma_F(g_*) \equiv \gamma,
\]
with $\beta(g_*)\simeq 0$ in this regime.
Under this standard RG assumption, Eq.~\eqref{eq:CS_freq} reduces, in the infrared, to
the linearised form
\begin{equation}
\left(
\omega\,\frac{d}{d\omega} - \gamma
\right)
\mathcal{F}_{\text{IR}}(\omega) \simeq 0,
\end{equation}
whose general solution is the power law
\begin{equation}
\mathcal{F}_{\text{IR}}(\omega)
= C \left( \frac{\omega}{\mu_0} \right)^{-\gamma},
\end{equation}
where $C$ is a dimensionless constant and $\mu_0$ is a reference scale.

Thus, the correction to the effective temperature arising from Berry--phase
effects follows from the infrared contribution
\begin{equation}
\Delta u_{\rm Berry}(\omega,T)
\simeq
-\frac{\hbar\,C}{2\pi^2 c^3}\,
\omega^{\,3-\gamma}+ {\cal O}(\omega, T).
\label{eq:Delta_u_Berry_IR}
\end{equation}

This expression shows that the Berry--phase correction induces a power--law
modification of the Rayleigh--Jeans law, controlled by the effective infrared
exponent~$\gamma$.
In observational terms, the corresponding deviation of the spectral energy
density scales as $\Delta u_{\rm Berry}(\omega)\propto \omega^{\,3-\gamma}$,
so that the effective critical index governing the Rayleigh--Jeans behaviour
is $(3-\gamma)$.

This correction to the Rayleigh--Jeans law, derived from the adiabatic approximation
of QED, shows good agreement with the observational data, as will be discussed in the
next section.

\section{ confronting the observational data from ARCADE2 measurements}

With the arguments of the previous section, the temperature excess to linear order is
\begin{equation}
\Delta T(\omega)
\simeq
-\,C\,
\frac{\hbar\,\omega_0^{\gamma}}{2k_{\rm B}}\,
\omega^{\,1-\gamma},
\label{eq:DeltaT_with_C}
\end{equation}
{The Berry-induced correction modifies the frequency dependence of the Rayleigh--Jeans spectrum but does not introduce inverse powers of the temperature. The leading Rayleigh--Jeans contribution remains linear in $T$, while the effective temperature excess $\Delta T(\omega)$ is defined by matching the corrected spectrum to the Rayleigh--Jeans form and is therefore temperature independent at leading order.}
\noindent
which represents a geometric correction induced by the Berry phase emerging in the infrared regime of quantum electrodynamics (IR-QED). 

Since the normalization factor $C$ can be adjusted to reproduce the measured excess, the expression is fully compatible with the observations at this level. However, determining the scaling exponent $\gamma$ is a more subtle and nontrivial problem.

In this framework, the exponent $\gamma$ appearing in the solution
\[
F_{\rm eff}(\omega)\sim \omega^{-\gamma}
\]
of the renormalization--group equation controls the infrared scaling of the Berry phase and therefore determines the Rayleigh--Jeans scaling symmetry through the effective factor $F_{\rm eff}$. Operationally, $\gamma$ organizes the Berry corrections in the infrared: the RG solution resums the iterative Berry contributions into a single power law. Although higher--order corrections could in principle be obtained iteratively, the present formulation does not provide a numerical prediction for~$\gamma$.

Between 2008 and 2011, the \textit{ARCADE~2} mission---a stratospheric balloon experiment developed by NASA---provided high--precision measurements of the absolute CMB spectrum in the low--frequency range \cite{Seiffert2011,Fixsen2011} It showed that, for frequencies below $10~\text{GHz}$, an excess temperature is observed that can be fitted as
\[
\Delta T(\omega)
\simeq
1.26~{\rm K}\,
\left(\frac{\omega}{1~{\rm GHz}}\right)^{-2.6}.
\]
Accordingly with these measurements our $\gamma$ coefficient should be 
\[
\gamma \simeq 3.6.
\]

The frequency $1~\text{GHz}$ does not define an internal physical scale; rather, it serves as a normalization reference that makes the ratio dimensionless. In other words, $1~\text{GHz}$ is the reference frequency around which the excess is measured, corresponding to $T\simeq1.26~\text{K}$.

The interpretation proposed by the \textit{ARCADE~2} collaboration--and discussed in other works such as \cite{Seiffert2011,Fixsen2011,Subrahmanyan2013,Condon2012}--is that the measured excess temperature arises from external astrophysical sources outside the CMB, whose origin remains uncertain. 

Possible explanations include synchrotron emission from faint, unresolved radio galaxies, or exotic processes such as the annihilation or decay of dark matter particles \cite{recent1,recent2,recent3}.

\begin{figure}[h!]
    \centering
    \includegraphics[width=0.95\linewidth]{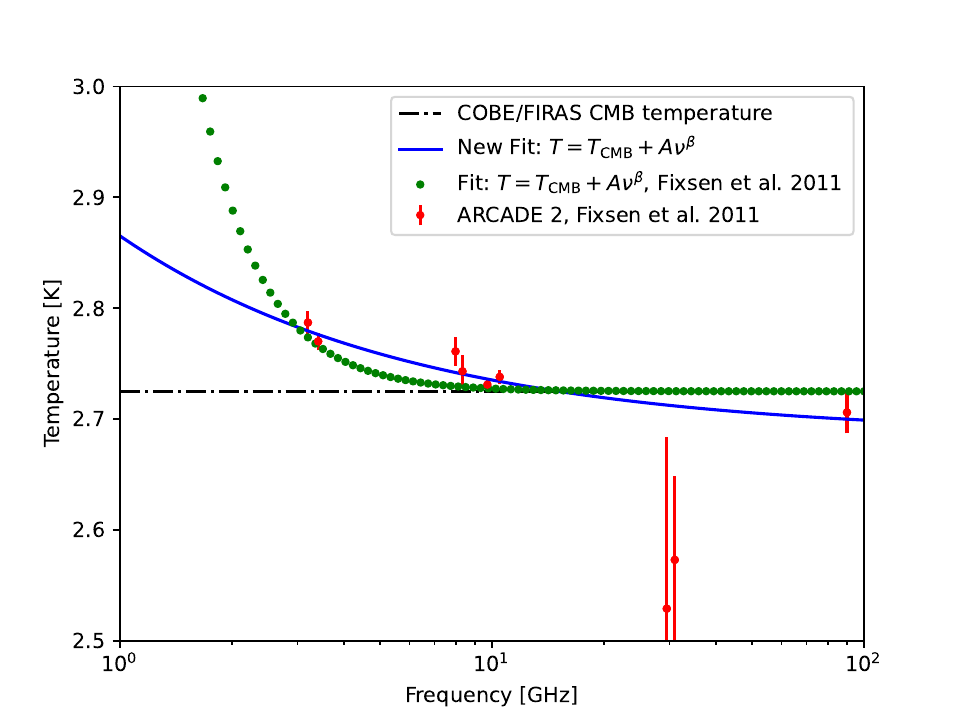}
    
    \caption{Thermodynamic temperature ($K$) as a function of frequency ($GHz$) for ARCADE 2 (in red), the vertical lines indicate $1 \sigma$ as listed in \cite{Fixsen2011}. The solid blue line represents the best fit to the ARCADE2 data, incorporating a constant CMB temperature and a power law component with an assumed index of $\beta$ and prefactor $A$.  The black dash dotted line shows the FIRAS CMB temperature \cite{2002ApJ...581..817F}. The green dotted line represents the best fit as chosen in \cite{Fixsen2011}.}
    \label{fig:arcade_fit}
\end{figure}

\Cref{fig:arcade_fit} shows the ARCADE~2 data as presented in \cite{Fixsen2011}.  
The $5\sigma$ excess at $3$~GHz above the FIRAS CMB value \cite{2002ApJ...581..817F} is shown together with the fit using the assumed spectral index $-2.6$ (green curve).  
This green fit follows the parameter choice of \cite{Fixsen2011}, where the excess is modeled as
\[
T(\omega)=T_{\rm CMB}+A\,\omega^{\beta},
\]
with the fitted values $T_{\rm CMB}=2.725~\mathrm{K}$, $A=1.26~\mathrm{K}$, and $\beta=-2.6$.  

Using a non-linear least-squares algorithm, we obtain slightly different best-fit parameters, plotted as the blue solid curve in \Cref{fig:arcade_fit}.  
Our fit yields
\[
T_{\rm CMB}=2.68508~\mathrm{K},\qquad
A=0.18005~\mathrm{K},\qquad
\beta=-0.555.
\]

Future experiments, such as the Primordial Inflation Explorer (PIXIE) \cite{pixie}, aim to measure the spectral energy distribution and linear polarization of the CMB and astrophysical foregrounds across $28 - 6000\, GHz$. This will enhance understanding of the findings from ARCADE 2 at $29.5$ and $31$ GHz. 

\section{Discussion of the results}

The analysis presented in this work shows that the infrared sector of QED naturally 
acquires a scale-invariant structure once the soft photon is treated as an 
electron--photon cloud transported adiabatically in the space of gauge configurations.  
In this framework the Berry connection becomes the relevant geometrical object, and 
its contribution to the energy density is encoded in the effective factor 
$F_{\rm eff}(\omega)$ that multiplies the Rayleigh--Jeans spectrum.  
The infrared scaling symmetry of the Rayleigh--Jeans law implies that $F_{\rm eff}$ 
must satisfy a renormalization--group equation whose solution is a power law 
$F_{\rm eff}(\omega)\sim\omega^{-\gamma}$.  
The exponent $\gamma$ plays the role of an effective critical exponent---or 
anomalous dimension---characterizing the infrared behaviour of the Berry cloud.

Once this scaling structure is assumed, the Berry correction to the energy density 
and the corresponding temperature excess follow immediately.  
To leading order, the excess obeys 
$\Delta T(\omega)\propto\omega^{\,1-\gamma}$, which is precisely the type of 
power--law behaviour inferred from the ARCADE~2 radio excess \cite{Seiffert2011,Fixsen2011,Subrahmanyan2013,Condon2012}.  
Matching the observed slope determines the value of the exponent, 
$\gamma\simeq 3.6$.  

At this stage, however, the numerical value of $\gamma$ is not predicted internally 
by IR--QED: the present construction identifies the scaling form and its geometrical 
origin in the Berry phase, but the exponent itself must be fixed phenomenologically 
from observational data.  

In this sense, $\gamma$ plays a role analogous to the early appearance of $h$ in 
Planck’s blackbody theory: it is a structural parameter whose full theoretical 
derivation requires a deeper dynamical understanding of the functional Berry 
connection.

Despite this limitation, the overall picture is conceptually coherent and 
remarkably simple: the infrared behaviour of QED is governed by adiabatically 
transported electron--photon clouds whose Berry holonomy produces an effective 
temperature shift in the radio domain.  
The observed ARCADE~2 excess \cite{Seiffert2011,Fixsen2011,Subrahmanyan2013,Condon2012} naturally within this framework, suggesting that 
infrared QED may have measurable consequences in cosmological radio observations.  
Further work is needed to derive $\gamma$ from first principles and to investigate 
possible signatures at lower frequencies or in complementary observables.

\acknowledgments
\noindent
The authors thank Rolando D{\"u}nner for helpful discussions. One of us (J.G.) would also like to express his gratitude to the members of the Quantum Friday–USACH seminar for their incisive and stimulating questions. This research was supported by DICYT (USACH), grant number 042531GR\_REG.
The work of N.T.A is supported by Agnes Scott College.


\begin{thebibliography}{99}
\bibitem{BN}
F. Bloch and A. Nordsieck, ``Note on the Radiation Field of the Electron'', \emph{Phys. Rev.} \textbf{52}, 54 (1937).



\bibitem{KI}
T. Kinoshita, ``Mass Singularities of Feynman Amplitudes'', \emph{J. Math. Phys.} \textbf{3}, 650 (1962).

\bibitem{LN}
T. D. Lee and M. Nauenberg, ``Degenerate Systems and Mass Singularities'', \emph{Phys. Rev.} \textbf{133}, B1549 (1964).

\bibitem{YFS}
D. R. Yennie, S. C. Frautschi, and H. Suura, ``The infrared divergence phenomena and high-energy processes'', \emph{Annals Phys.} \textbf{13}, 379 (1961).

\bibitem{Weinberg} S.~Weinberg,
``Infrared photons and gravitons,''
Phys. Rev. \textbf{140} (1965), B516-B524
doi:10.1103/PhysRev.140.B516.

\bibitem{chung} See also V.~Chung,
``Infrared Divergence in Quantum Electrodynamics,''
Phys. Rev. \textbf{140} (1965), B1110-B1122
doi:10.1103/PhysRev.140.B1110.

\bibitem{kibble1} T.~W.~B.~Kibble,
``Coherent Soft-Photon States and Infrared Divergences. I. Classical Currents,''
J. Math. Phys. \textbf{9} (1968) no.2, 315-324.

\bibitem{kibble2} T.~W.~B.~Kibble, ``Coherent soft-photon states and infrared divergences. ii. mass-shell singularities of green's functions,''
Phys. Rev. \textbf{173} (1968), 1527-1535
doi:10.1103/PhysRev.173.1527.

\bibitem{kibble3} T.~W.~B.~Kibble, ``Coherent soft-photon states and infrared divergences. iii. asymptotic states and reduction formulas,''
Phys. Rev. \textbf{174} (1968), 1882-1901
doi:10.1103/PhysRev.174.1882.

\bibitem{kibble4} T.~W.~B.~Kibble, ``Coherent soft-photon states and infrared divergences. iii. asymptotic states and reduction formulas,''
Phys. Rev. \textbf{174} (1968), 1882-1901
doi:10.1103/PhysRev.174.1882.
Phys. Rev. \textbf{175} (1968).

\bibitem{KF}
P. P. Kulish and L. D. Faddeev, ``Asymptotic conditions and infrared divergences in quantum electrodynamics'', \emph{Theor. Math. Phys.} \textbf{4}, 745 (1970).

\bibitem{sorella} M.~S.~Guimaraes, I.~Roditi, S.~P.~Sorella and A.~F.~Vieira,
``A numerical analysis of Araki-Uhlmann relative entropy in Quantum Field Theory,''
Nucl. Phys. B \textbf{1018} (2025), 117011
doi:10.1016/j.nuclphysb.2025.117011.

\bibitem{araki} H. Araki, Publ. Res. Inst. Math. Sci. Kyoto 1976, 809 (1976).
 \bibitem{araki1} A.~Uhlmann,
``Relative Entropy and the Wigner-Yanase-Dyson-Lieb Concavity in an Interpolation Theory,''
Commun. Math. Phys. \textbf{54} (1977), 21
doi:10.1007/BF01609834 

\bibitem{JG1} J.~Gamboa,
``Entanglement and effective field theories,''
Phys. Lett. B \textbf{868} (2025), 139723
doi:10.1016/j.physletb.2025.139723
[arXiv:2502.11819 [hep-th]].

\bibitem{JG2} J.~Gamboa,
``Topology and the infrared structure of quantum electrodynamics,''
JHEP \textbf{07} (2025), 184
doi:10.1007/JHEP07(2025)184
[arXiv:2505.13247 [hep-th]].

\bibitem{JG3} J.~Gamboa and F.~Mendez,
``QED-IR as topological quantum theory of dressed states,''
JHEP \textbf{11} (2025), 040
doi:10.1007/JHEP11(2025)040
[arXiv:2507.11668 [hep-ph]].

\bibitem{CGL}
J.~L.~Cortés, J.~Gamboa, S.~Lepe and J.~López-Sarrión,
``An Adiabatic approximation to the path integral for relativistic fermionic fields,''
Phys. Lett. B \textbf{619} (2005), 367-376
doi:10.1016/j.physletb.2005.04.044
[arXiv:hep-ph/0204054 [hep-ph]].


\bibitem{Berry2}
M.~V.~Berry,
``Quantum Phase Corrections from Adiabatic Iteration'',
Proc. Roy. Soc. Lond. A \textbf{414} (1987), 31-46.


\bibitem{Fixsen2011}
D.~J.~Fixsen, A.~Kogut, S.~M.~Levin, M.~Limon, P.~Mirel, S.~Seiffert, 
J.~Singal, T.~Villela, C.~A.~Wollack and E.~J.~Wollack,
``ARCADE 2 Measurement of the Absolute Sky Brightness at 3--90 GHz,''
\textit{Astrophys.\ J.} \textbf{734}, 5 (2011).
doi:10.1088/0004-637X/734/1/5

\bibitem{Seiffert2011}
M.~Seiffert, D.~J.~Fixsen, A.~Kogut, J.~C.~Mather, S.~H.~Moseley, 
E.~J.~Wollack and M.~Limon,
``Interpretation of the ARCADE 2 Absolute Sky Brightness Measurement,''
\textit{Astrophys.\ J.} \textbf{734}, 6 (2011).
doi:10.1088/0004-637X/734/1/6

\bibitem{Condon2012}
J.~J.~Condon, W.~D.~Cotton, E.~B.~Fomalont, K.~I.~Kellermann, N.~Miller, 
R.~A.~Perley and N.~W.~B.~Benson,
``Resolving the Radio Source Background: Deeper Understanding of the ARCADE 2 Excess,''
\textit{Astrophys.\ J.} \textbf{758}, 23 (2012).
doi:10.1088/0004-637X/758/1/23

\bibitem{Subrahmanyan2013}
R.~Subrahmanyan and R.~D.~Ekers,
``The Cosmic Radio Background and Implications for the ARCADE 2 Excess,''
\textit{Mon.\ Not.\ Roy.\ Astron.\ Soc.} \textbf{434}, 293--303 (2013).
doi:10.1093/mnras/stt1007

\bibitem{recent1} B.~Dev, P.~Di Bari, I.~Martinez-Soler and R.~Roshan,
``Boomerang mechanism explaining the excess radio background,''
[arXiv:2509.03441 [hep-ph]].
\bibitem{recent2} J.~Singal, J.~Haider, M.~Ajello, D.~R.~Ballantyne, E.~Bunn, J.~Condon, J.~Dowell, D.~Fixsen, N.~Fornengo and B.~Harms, \textit{et al.}
%``The Radio Synchrotron Background: Conference Summary and Report,''
Publ. Astron. Soc. Pac. \textbf{130} (2018) no.985, 036001
doi:10.1088/1538-3873/aaa6b0
[arXiv:1711.09979 [astro-ph.HE]].

\bibitem{recent3} M.~Ruchika, W.~Giar{\`e}, E.~M.~Teixeira and A.~Melchiorri,
%``Resilience and implications of adiabatic CMB cooling,''
Phys. Dark Univ. \textbf{49} (2025), 101999
doi:10.1016/j.dark.2025.101999
[arXiv:2505.02909 [astro-ph.CO]].

\bibitem{2002ApJ...581..817F}Fixsen, D. \& Mather, J. The Spectral Results of the Far-Infrared Absolute Spectrophotometer Instrument on COBE. {\em \apj}. \textbf{581}, 817-822 

\bibitem{pixie}
A.~Kogut, N.~Aghanim, J.~Chluba, D.~T.~Chuss, J.~Delabrouille, C.~Dvorkin, D.~Fixsen, S.~Ghosh, B.~S.~Hensley and J.~C.~Hill, \textit{et al.}
``The Primordial Inflation Explorer (PIXIE): mission~design and science goals,''
JCAP \textbf{04} (2025), 020
doi:10.1088/1475-7516/2025/04/020


\end{thebibliography}
\end{document}